\documentclass[]{raa}
\usepackage{graphicx}
\usepackage{times}
\usepackage{natbib}
\usepackage{mathrsfs}
\usepackage{amsmath}
\usepackage{amssymb}
\usepackage{booktabs}
\usepackage{hyperref}
\def\etal {et al.~}

\newcommand{\choh}{CH$_3$OH }                             

\begin{document}

\title{A 95 GHz Methanol Emission Survey Toward Eight Small Supernova Remnants}

\volnopage{Vol.0 (200x) No.0, 000--000}

   \setcounter{page}{1}

\author{Yingjie Li\inst{1, 2}\and Ye Xu\inst{1}\and Xi Chen\inst{3, 4}\and Deng-Rong Lu\inst{1}\and Yan Sun\inst{1}\and Xinyu Du\inst{1, 5}\and Zhi-Qiang Shen\inst{4}}

\institute{Purple Mountain Observatory, Chinese Academy of Sciences, Nanjing 210008, China; {\it xuye@pmo.ac.cn}\\
\and
University of Science and Technology of China, Chinese Academy of Sciences, Hefei, Anhui 230026, China; {\it liyj@pmo.ac.cn}\\
\and
Center for Astrophysics, GuangZhou University, Guangzhou 51006, China; {\it chenxi@shao.ac.cn}\\
\and
Shanghai Astronomical Observatory, Chinese Academy of Sciences, Shanghai 200030, China\\
\and
Graduate University of the Chinese Academy of Sciences, 19A Yuquan Road, Shijingshan District, Beijing 100049, China\\
}

\abstract{We report on a 95 GHz ($8_0-7_1$ A$^{+}$) methanol (CH$_3$OH) emission survey with the Purple Mountain Observatory Delingha 13.7 m telescope. Eight supernova remnants (SNRs) with angular size $\lesssim$ 10$\arcmin$ were observed, but emission was only detected in three SNRs near the Galactic center (Sgr A East, G 0.1-0.1, and G 359.92-0.09). \choh emission mainly surrounds the SNRs and can be decomposed into nine spatial peaks with velocity range of eight peaks being (-30, 70) km s$^{-1}$, and the other (70, 120) km s$^{-1}$. They are probably excited by interaction with these SNRs and adjacent molecular gas in the central molecular zone (CMZ), although star formation may play an important role in exciting \choh emission in some regions of CMZ. We infer that tidal action is unlikely to be an excitation source for \choh emission.
}

\authorrunning{Y. Li et al.}

\titlerunning{95 GHz Methenol Emission Toward SNRs}
\maketitle

\section{Introduction}

The methanol molecule (CH$_3$OH) is a slightly asymmetric top with hindered internal rotation. It processes a multitude of allowed radio transitions. \choh in space has been actively studied since its discovery by \citet{BGL1970}, abundant \choh transitions have been detected at sub-millimeter and millimeter wavelengths, and are sensitives to the density and the kinetic temperature of the dense clouds \citep[see e.g., ][and references therein]{MWH1988, MRM1988, LSM2004}. \choh is one of the most abundant complex molecules in star formation regions \citep{KDB1997}, where both its thermal \citep{SBB1994, HTM1998} and maser (see below) lines can trace star formation activities. However, for \choh lines at 95 GHz ($8_0-7_1$ A$^{+}$), the attention has mostly been concentrated on its maser (belong to class I \choh maser lines) emission though there are a few detections of thermal emission as well \citep[see e.g., ][]{VES2000, CEH2012}.

Class I \choh maser lines that are pumped by collisions, such as at 36 GHz ($4_{-1}-3_0$ E), 44 GHz ($7_0-6_1$ A$^{+}$), 95 GHz ($8_0-7_1$ A$^{+}$), etc., have long been assumed to be associated with both low-mass and high-mass star formation \citep{KPS2006, CES2011, YXC2017}. From the beginning of 2011, targeted searches for these masers toward supernova remnants (SNRs) had started carrying out \citep{LAV2011}. These maser emission, such as at 36 GHz ($4_{-1}-3_0$ E) and 44 GHz ($7_0-6_1$ A$^{+}$), has been detected in some supernova remnants (SNRs) \citep{LAV2011, PSF2011, PSF2014}. However, there is only one detection of 95 GHz \choh maser line in SNRs \citep[i.e., SNR Kes 79,][]{ZS2008}, and it is controversial in that \citet{F2011} claimed that the 12 m Arizona Radio Observatory failed to confirm this detection. The \choh maser line at 95 GHz is excited under a stricter range of conditions than that at 36 and 44 GHz \citep[see e.g.,][]{MPS2014, N2016}. That may be the factor that fewer 95 GHz \choh maser emission was detected in SNRs. The factor may also be an effect of not having a large survey of SNRs at 95 GHz. We interest in whether \choh emission is associated with SNRs.

\citet{YCW2013} explained the enhanced abundance of \choh in the Galactic center in terms of the interaction between cosmic rays and molecular gas, which produces $\gamma$-rays \citep{F2011}. Therefore, the presence of $\gamma$-ray emission such as at Gev or Tev energies marks a possibility of enhanced abundance of \choh (such as at 36, 44 and 95 GHz), and hence may raise the possibility of detection. Although the excitation conditions for these three masers are different, all three maser emission outputs are presented over a wide range of environments \citep[see e.g.,][]{MPS2014, LMW2016}. In order to expect expect a high detection rate and save time, the selection criteria were that the SNRs were small ($\lesssim 10\arcmin$) and associated with class I methanol 36/44 GHz maser or GeV/TeV emission. We selected eight SNRs, with three (Sgr A East, G 0.1-0.1, and G 359.92-0.09) located in the Galactic center. Table 1 summarizes the physical properties of the selected SNRs, including their interaction with clouds, occurrence of 1720 MHz OH and class I methanol 36/44 GHz maser, SNR type, and high-energy emission.


\begin{table}
\bc
\begin{minipage}[]{100mm}
\caption[]{SNR properties}\end{minipage}
\scriptsize
\begin{tabular}{lccccccc}
\hline\hline
Name & Interacting with & OH 1720 MHz & \choh 36/44 & Type & X-ray & GeV Flux$^\mathrm{c}$ & TeV$^\mathrm{2}$ ($\sigma$) \\
   & molecular clouds$^\mathrm{1}$ & maser$^\mathrm{a}$ & GHz maser$^\mathrm{b}$ &  &  & ($\times10^{-9}$ ph cm$^{-2}$ s$^{-1}$) & \\
\hline
Sgr A East & Y & Y & Y & TC$^\mathrm{d}$ & Y$^\mathrm{e}$ & 65.02 & HESS(66.6)  \\
G 0.1-0.1 & Y? &   &   & C?$^\mathrm{f}$ & Y$^\mathrm{g}$ & 1.0$^\mathrm{3}$ & HESS (11.5) \\
G 1.4-0.1 & Y & Y & Y  & S$^\mathrm{h}$  &                    & 0.4$^\mathrm{3}$ & HESS (6.5) \\
G 29.7-0.3 & Y & Y & N & C$^\mathrm{i}$ & Y$^\mathrm{j}$ & 2.1$^\mathrm{3}$ & HESS (10.1)\\
G 111.7-2.1 & Y? & N & & S$^\mathrm{k}$ & Y$^\mathrm{l}$ & 6.25& MAGIC (5.2)  \\
G 120.1+1.4 & Y? & N & & S$^\mathrm{m}$ & Y$^\mathrm{l}$ & 2.1$^\mathrm{3}$ & VERITAS (5.8) \\
G 184.6-5.8 & ?  &   & & F$^\mathrm{n}$ & Y$^\mathrm{l}$ & 108.12 &  HESS (129) \\
G 359.92-0.09$^\mathrm{4}$ & Y$^\mathrm{o}$ &  & & &  &  & \\
\hline
 \end{tabular}
 \ec
\tablerefs{1.0\textwidth}{(a)~\citet{FGR1996, KFG1998, YGR1999, DOK2009};
(b) \citet{LAV2011, PSF2011, PSF2014};
(c) \citet{AAA2016};
(d) \citet{YWR2003, V2012};
(e) \citet{MBF2002};
(f) \citet{HW2013b};
(g) \citet{YLWe2002};
(h) \citet{F2011};
(i) \citet{BHS1983, BH1996};
(j) \citet{SMK2001};
(k) \citet{AAB2001, AAA2007};
(l) \citet{PS1995};
(m) \citet{AAA2011};
(n) \citet{WLT1994, WLK1999};
(o) \citet{CH2000, HH2005, AMM2011}.
}
\tablecomments{1.0\textwidth}{
In the column of ``Type'', S=shell, C=composite, F=plerion, TC=thermal composite (i.e. mixed-morphology), C?=probably thermal \& plerionic composite. In the other columns, Y=yes, N=no, Y?=probable, ?=possible, Null=no data.\\
(1) see ``A List of Galactic SNRs Interacting with Molecular Clouds'', and references therein (\url{http://astronomy.nju.edu.cn/~ygchen/others/bjiang/interSNR6.htm})  .\\
(2) This column includes TeV $\gamma$-ray detections by experiment and significance of detection in parentheses for HESS \citep{AABe2006, AAB2008, B2011}, MAGIC \citep{AAA2007} and VERITAS \citep{AAA2011}.\\
(3) Upper limits of flux with 99\% confidence and PL index $\Gamma=2$, see Table 3 in \citet{AAA2016}.\\
(4) This SNR is included neither in ``A census for high-energy-observations of Galactic supernova remnants'' \citep[\url{http://www.physics.umanitoba.ca/snr/SNRcat/},][]{FS2012} nor in ``A catalogue of Galactic supernova remnants''\citep[\url{http://www.mrao.cam.ac.uk/surveys/snrs/},][]{G2014}.\\
}

\end{table}

We report on a 95 GHz \choh emission survey toward eight SNRs using the Purple Mountain Observatory Delingha (PMODLH) 13.7 m radio telescope. Observation and data reduction are described in Section 2, and the survey outcomes in Section 3. Section 4 discusses whether the 95 GHz \choh emission can be identified as maser lines. In section 5, we discuss the correlation of 95 GHz \choh emission with SNRs, star formation and tidal action, and why 95 GHz \choh emission was only detected around Sgr A East, G 0.1-0.1 and G 359.92-0.09. Finally, Section 6 summarizes the important results.

\section{Observation and Data Reduction}

Using the Purple Mountain Observatory Delingha (PMODLH) 13.7 m telescope, CH$_3$OH (8$_0$-7$_1$ A$^+$, 95.16949 GHz, hereafter 95 GHz CH$_3$OH) line was observed toward eight SNRs from May 5 to June 25, 2015. The telescope operates in sideband separation mode, and uses a fast Fourier transform spectrometer \citep{YS2012}.

Sgr A East, G 0.1-0.1 and G 359.92-0.04 were mapped in a single image spanning $\sim 20\arcmin\times20\arcmin$ (49.5 pc$^2$ at 8.5 kpc) cell (denote this region as Galactic central SNRs region, GCSNRR), and the remaining five SNRs were mapped in 10$\arcmin\times$10$\arcmin$ cells (see Table 2). We used the on-the-fly (OTF) observational mode with a constant rate of 50$\arcsec$ s$^{-1}$ along lines in R.A. and Decl. directions on the sky. The receiver recorded spectra every 0.3 s. The standard chopper wheel method was used for calibration \citep{PB1973, UH1976}. The half-power beam width of the telescope was approximately $58\arcsec$ at 95 GHz. Typical system temperature, $T_{\mathrm{sys}}$, during the observations was $160-190$ K. Antenna temperatures, $T^{\ast}_{\mathrm{A}}$, were calibrated to the main beam temperatures, $T^{\ast}_{\mathrm{R}}$, with main beam efficiency $\eta_{\mathrm{mb}}$ = 62\%. The calibrated data were re-gridded to $30\arcsec$ pixels. Table 2 summarizes mapping center and main beam root mean square noise (RMS) per 61 kHz channel (corresponding to 0.19 km s$^{-1}$ at 95 GHz) for each target source.

\begin{table}[!hbt]
\bc
\begin{minipage}[]{100mm}
\caption[]{Observed SNRs Positions and Observing RMS}\end{minipage}
\begin{tabular}{lcccc}          
\hline\hline
Region &  R.A. & Decl. & Size & RMS \\
 & (J2000) & (J2000) & (arcmin) & (mK) \\
\hline
 GCSNRR$^{\mathrm{1}}$ & 17 45 44.0 & -29 00 00 & 20 & 25 \\
 G 1.4-0.1 & 17 49 39.0 & -27 46 00 & 10 & 25 \\
 G 29.7-0.3 & 18 46 25.0 & -02 59 00 & 10 & 39 \\
 G 111.7-2.1 & 23 23 26.0 &  \ 58 48 00 & 10 & 28 \\
 G 120.1+1.4 & 00 25 18.0 &  \ 64 09 00 & 10 & 29 \\
 G 184.6-5.8 & 05 34 31.0 &  \ 22 01 00 & 10 & 27 \\
 \hline
  \end{tabular}
  \ec
\tablecomments{1.0\textwidth}{The columns show the name, mapping center in equatorial coordinates (J2000), mapping size and RMS per channel of each region.\\
(1) This region covers SNRs Sgr A East, G 0.1-0.1 and G 359.92-0.09.\\
}
\end{table}

\section{Results}

The survey resulted in three detections (Sgr A East, G 0.1-0.1 and G 359.92-0.09) that are covered by GCSNRR. We focused on the general distribution of 95 GHz \choh emission in GCSNRR in this section.

Figure 1 shows 95 GHz \choh integrated intensity map, where Sgr A$^{\ast}$ is indicated by a blue filled pentagram. The 95 GHz \choh emission south of Decl.$\sim$ -29$\degr$02$\arcmin$ shows two distinct velocity components. We integrated the emission separately over the low velocity component, -30 to 70 km s$^{-1}$ (Figure 1, the left panel), and high velocity component, 70 to 120 km s$^{-1}$ (Figure 1, the right panel).

Nine spatial peaks were evident, shown in Figure 1 by diamonds, and listed in Table 3. Only one position, located in the southwestern region of the map, contains two peaks, corresponding to peak 8 for the low velocity gas (Figure 1, the left panel) and peak 9 for the high velocity (Figure 1, the right panel). Peak 9, where velocity is centered at $\sim 85$ km s$^{-1}$, is much weaker with integral intensity of 0.47 K km s$^{-1}$. Figure 1 also indicates that most 95 GHz \choh emission surrounds SNRs. The correlation between them would be discussed in section 5.1.

\begin{figure*}[!ht]
\centering
\includegraphics[width=7.2cm,angle=0]{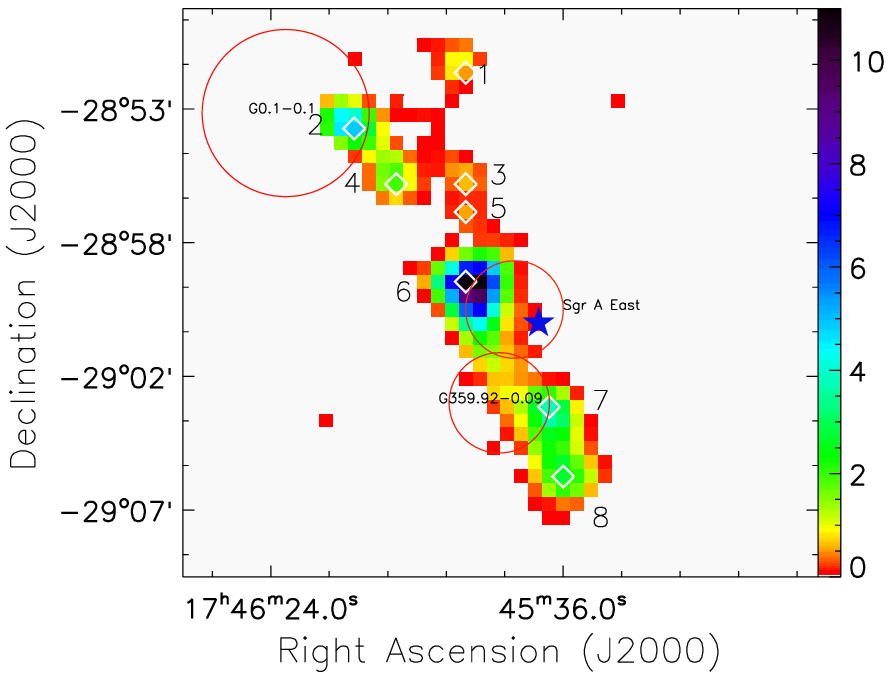}
\includegraphics[width=7.2cm,angle=0]{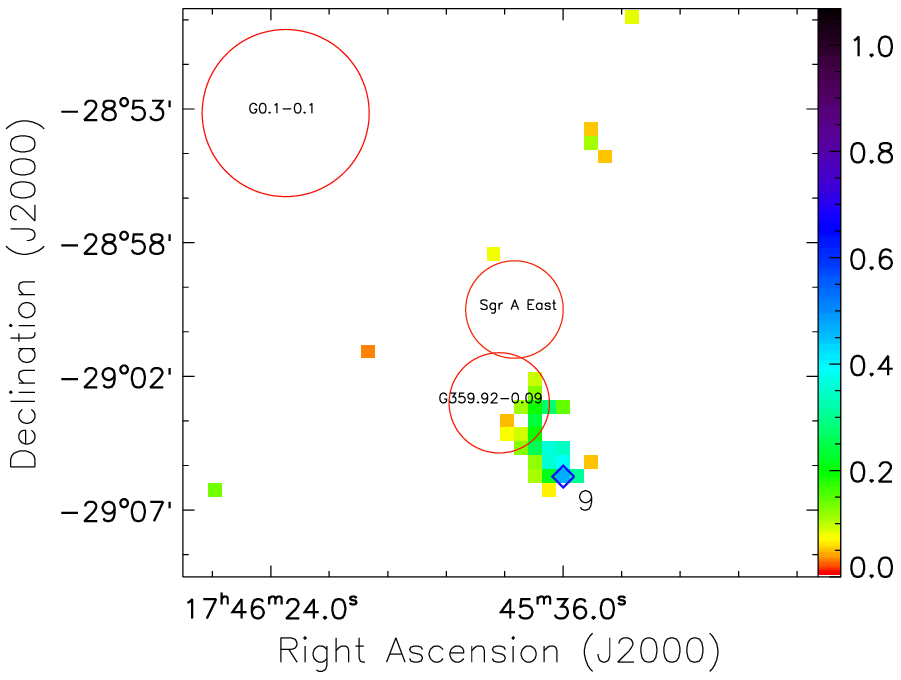}
\caption{Integrated intensity maps of 95 GHz CH$_3$OH, integrated over -30 to 70 km s$^{-1}$ (left), and 70 to 120 km s$^{-1}$ (right), respectively. Color bars units are K km s$^{-1}$. Maximum integral intensity of 95 GHz CH$_3$OH in the left and right panels are 10.94 and 0.47 K km s$^{-1}$, and the scales are 11.0 and 1.1 K km s$^{-1}$, respectively. The blue filled pentagram indicates the position of Sgr A$^{\ast}$. The diamonds show peak positions. The red circles indicate SNRs. The circles indicating G 359.92-0.09 and Sgr A East are the real angular size. The size of SNR G 0.1-0.1 is uncertain.}
\end{figure*}

Figure 2 presents the spectra of the nine emission peaks. The spectrum of peak 4 shows a bimodal structure. For peak 5, the spectrum shows a possible second velocity component centered at $\sim$ 45 km s$^{-1}$, where the peak $T_R^{\ast}$ of this component is less than 3 $\times$ RMS. The spectrum of peak 7 indicates a second velocity component centered at $\sim$ 85 km s$^{-1}$, which surrounds peak 9 from spatial distribution (see the right panel of Figure 1). The spectra of the remaining peaks are centrally dominated.

\begin{figure*}[!ht]
\centering
\includegraphics[width=4.0cm]{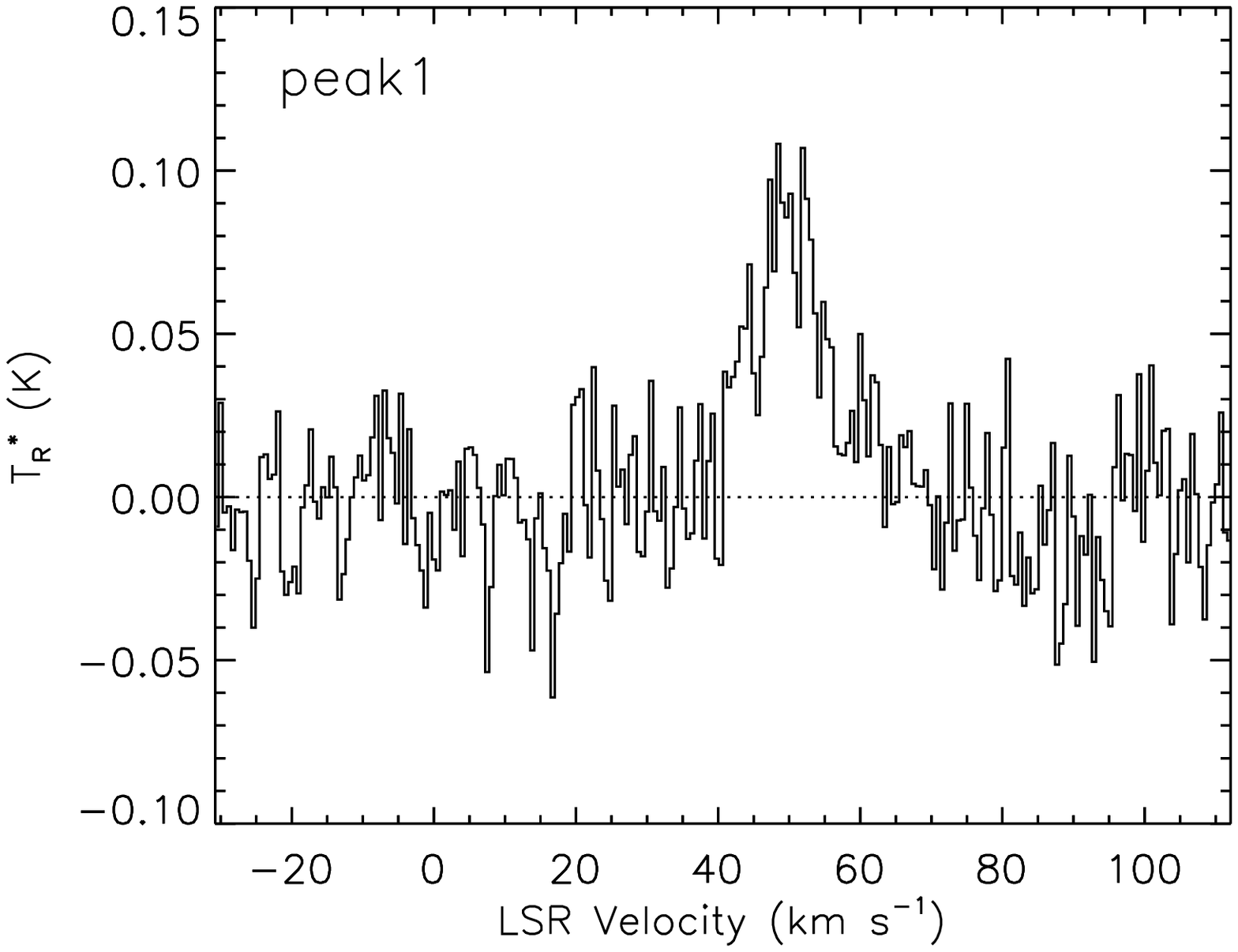}
\includegraphics[width=4.0cm]{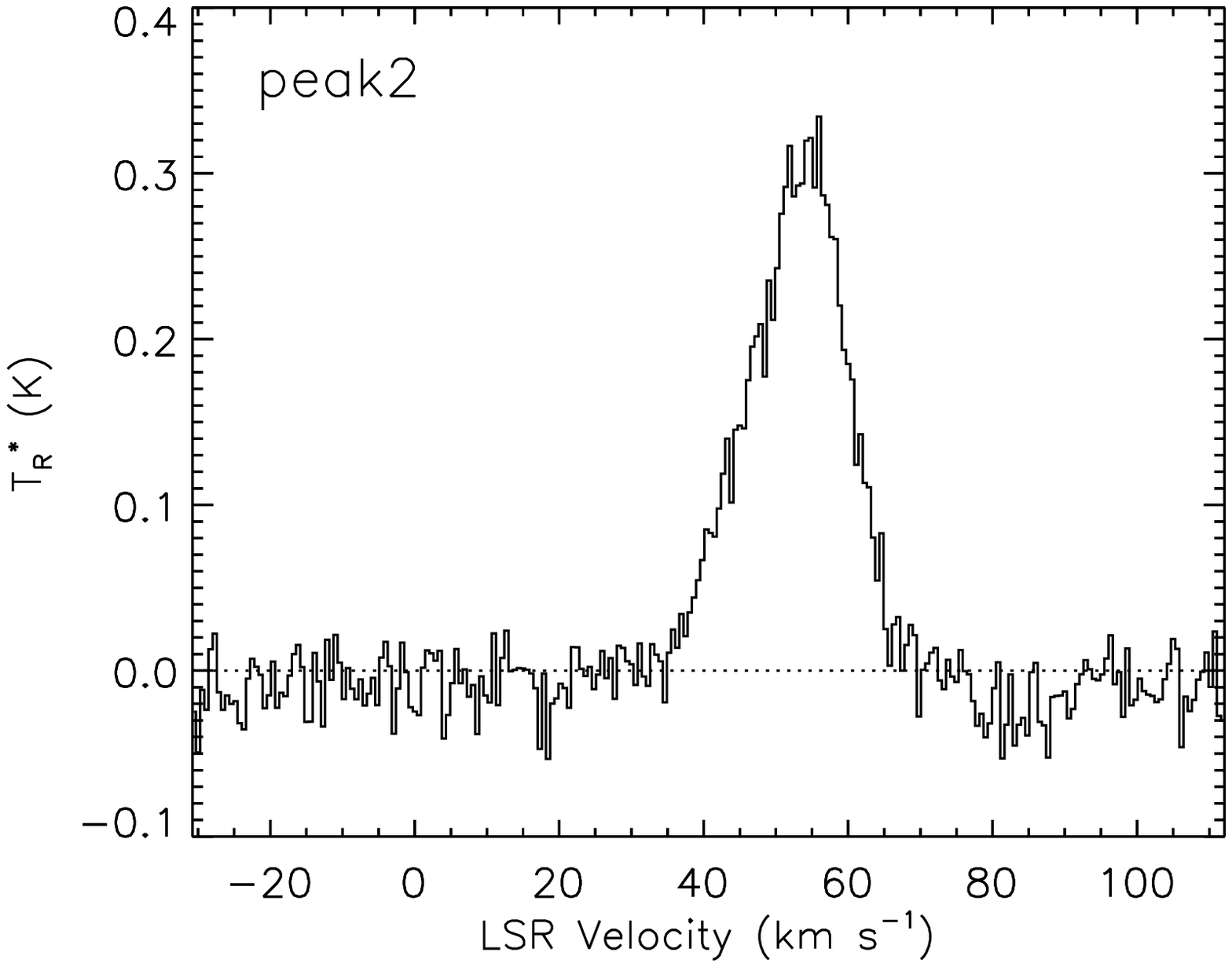}
\includegraphics[width=4.0cm]{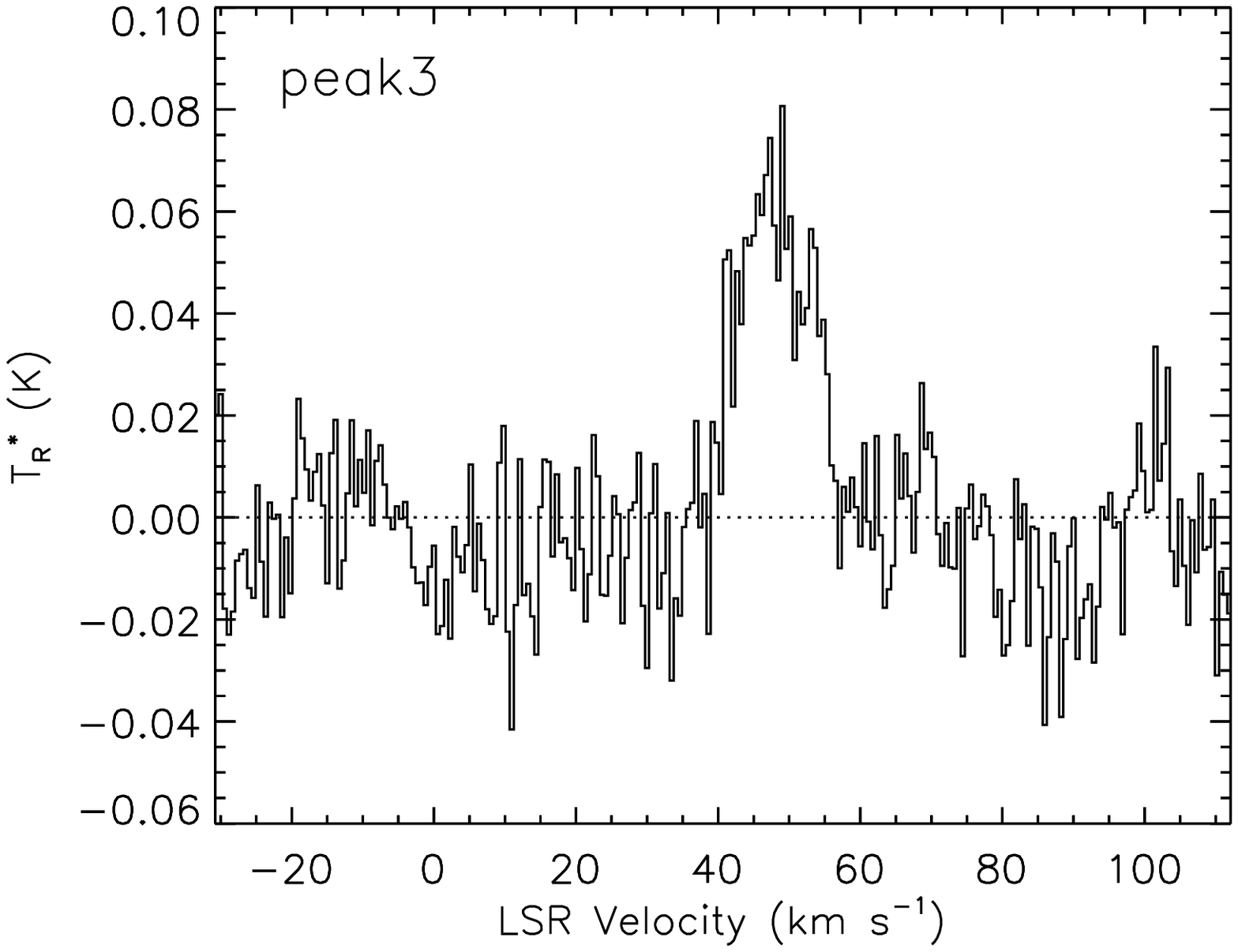}
\includegraphics[width=4.0cm]{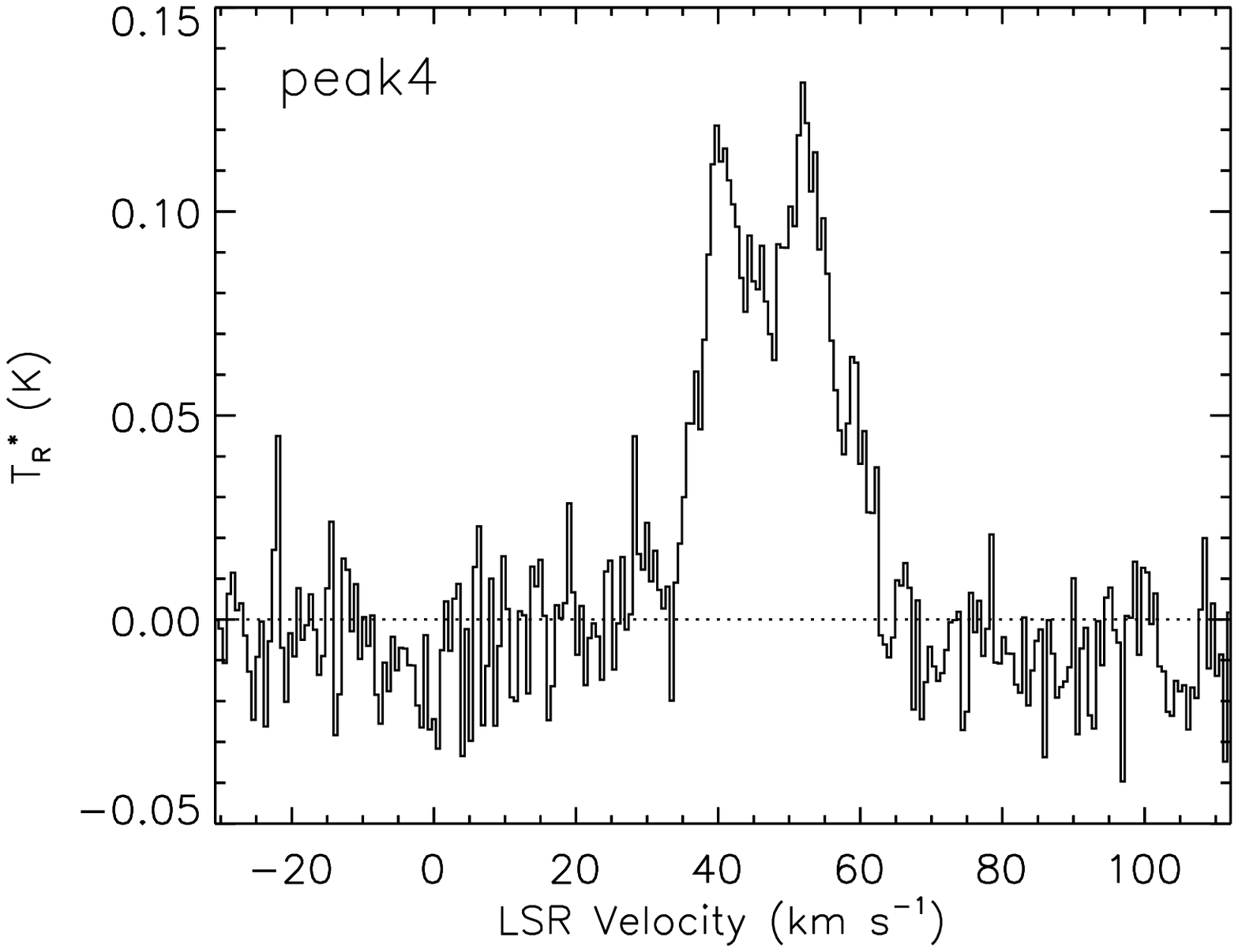}
\includegraphics[width=4.0cm]{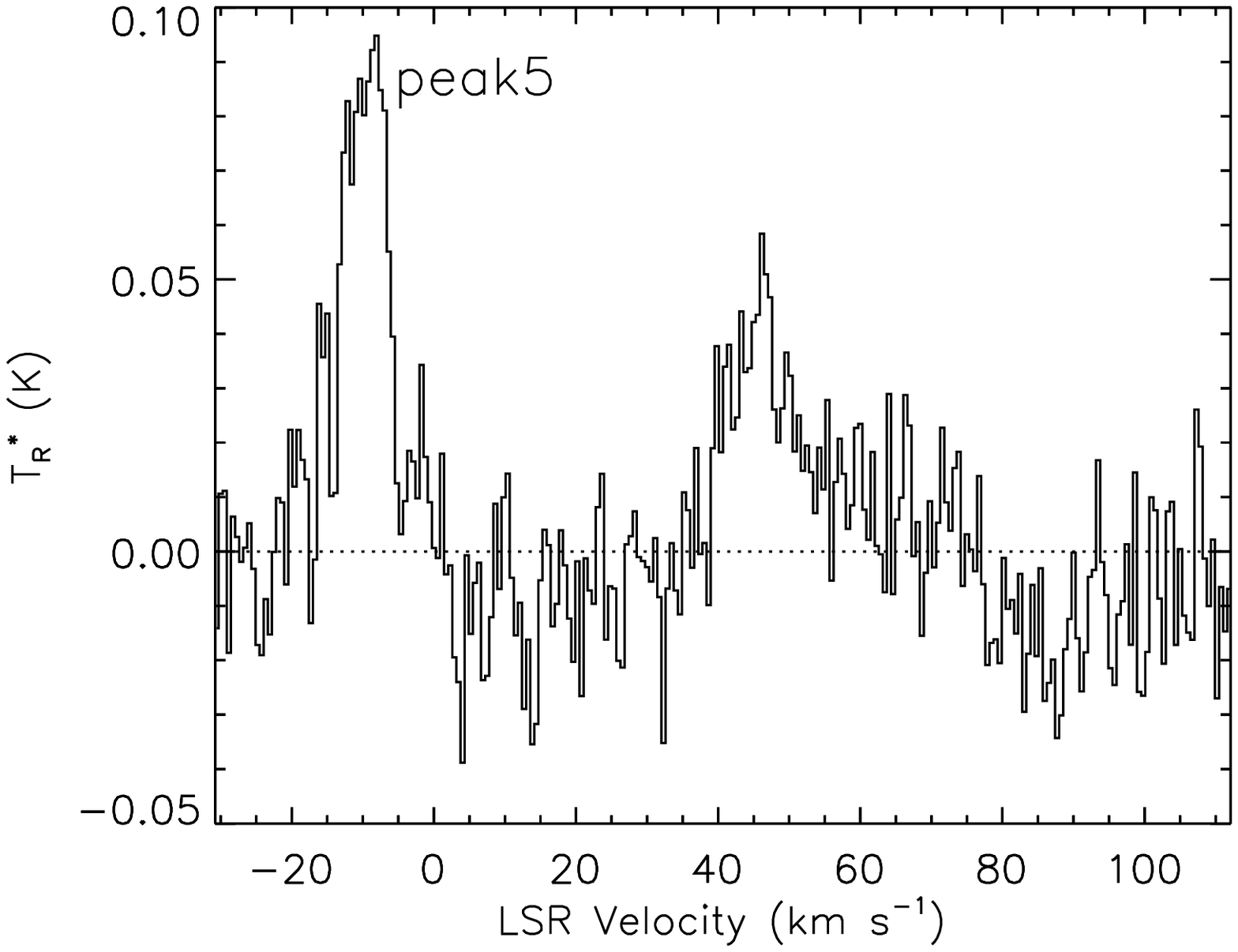}
\includegraphics[width=4.0cm]{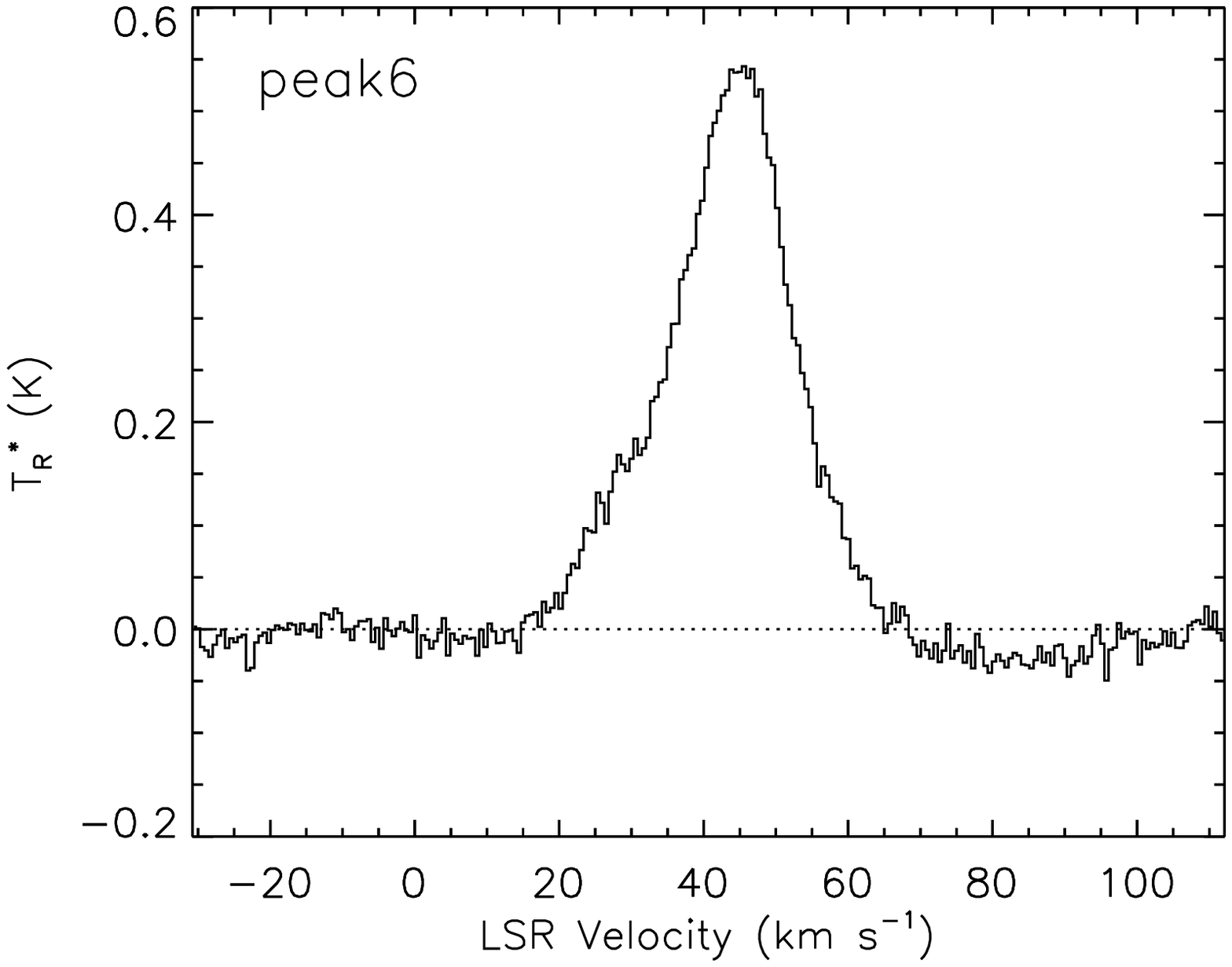}
\includegraphics[width=4.0cm]{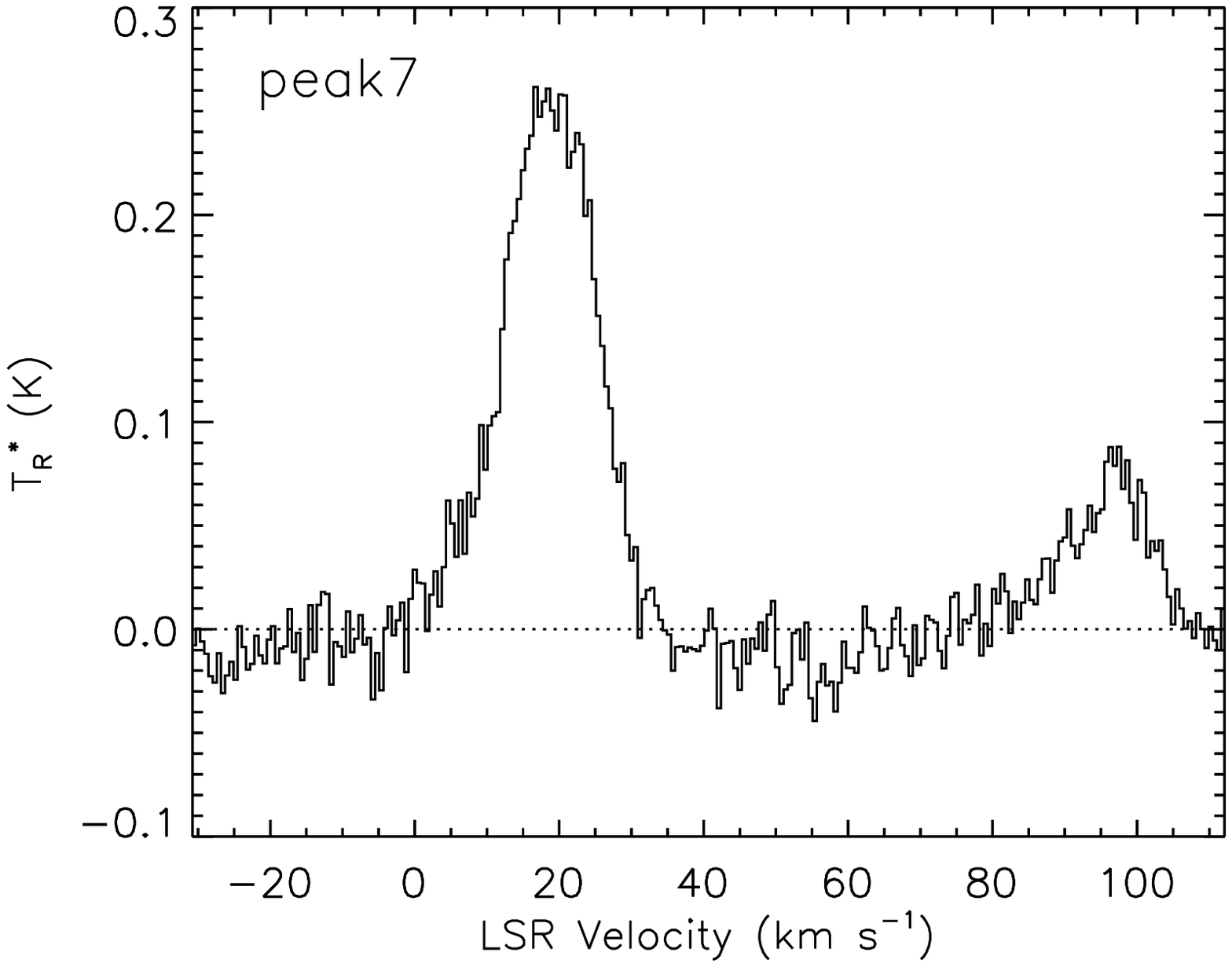}
\includegraphics[width=4.0cm]{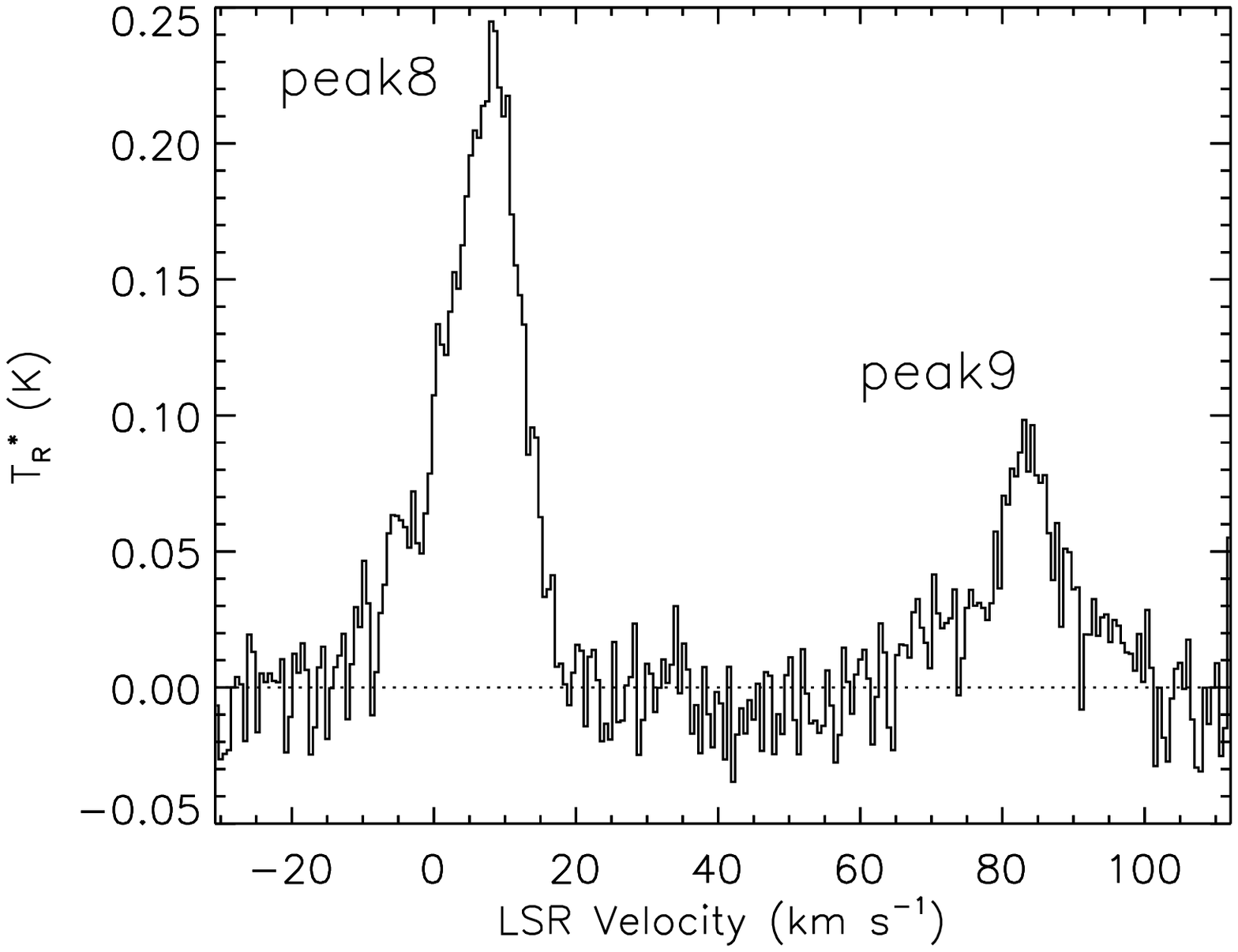}
\caption{Spectra of peaks from Table 3. Spectra are averaged over three velocity channels.}
\end{figure*}

\begin{table}[!hbt]
\bc
\begin{minipage}[]{100mm}
\caption[]{Peak Emission for 95 GHz \choh}\end{minipage}
\begin{tabular}{ccccc}
\hline\hline
Peak ID & R.A. & Decl. & Velocity & Integrated intensity\\
 & (J2000) & (J2000) & (km s$^{-1}$) & (K km s$^{-1}$)\\
\hline
  peak 1 & 17:45:52.0 & -28:51:30 & 49.9 & 0.92 \\
  peak 2 & 17:46:10.3 & -28:53:30 & 53.0 & 4.80 \\
  peak 3 & 17:45:52.0 & -28:55:30 & 46.7 & 0.59 \\
  peak 4 & 17:46:03.4 & -28:55:30 & 47.0 & 1.93 \\
  peak 5 & 17:45:52.0 & -28:56:30 & -6.4 & 0.52 \\
  peak 6 & 17:45:52.0 & -28:59:00 & 43.2 & 10.94 \\
  peak 7 & 17:45:38.3 & -29:03:30 & 18.8 & 3.73 \\
  peak 8 & 17:45:36.0 & -29:06:00 &  6.8 & 2.65 \\
  peak 9 & 17:45:36.0 & -29:06:00 & 84.8 & 0.47 \\
\hline
 \end{tabular}
 \ec
\tablecomments{1.0\textwidth}{The fourth column is the intensity weighted velocity (moment 1). The last column is integrated intensity which integrated over velocity from -30 to 70 km s$^{-1}$ for peaks 1-8, and from 70 to 120 km s$^{-1}$ for peak 9.
 }
\end{table}

Figure 3 shows the velocity (intensity weighted, i.e., moment 1) structure of the entire region for both the low (the left panel) and high velocity component (the right panel). A significant velocity gradient from the southwest to northeast is evident in both panels. Velocities for the low velocity component (the left panel) vary from $\sim$ 10 to $\sim$ 60 km s$^{-1}$, and from $\sim$ 80 to $\sim$ 100 km s$^{-1}$ for the high velocity component (the right panel).

\begin{figure*}[!ht]
\centering
\includegraphics[width=7.2cm,angle=0]{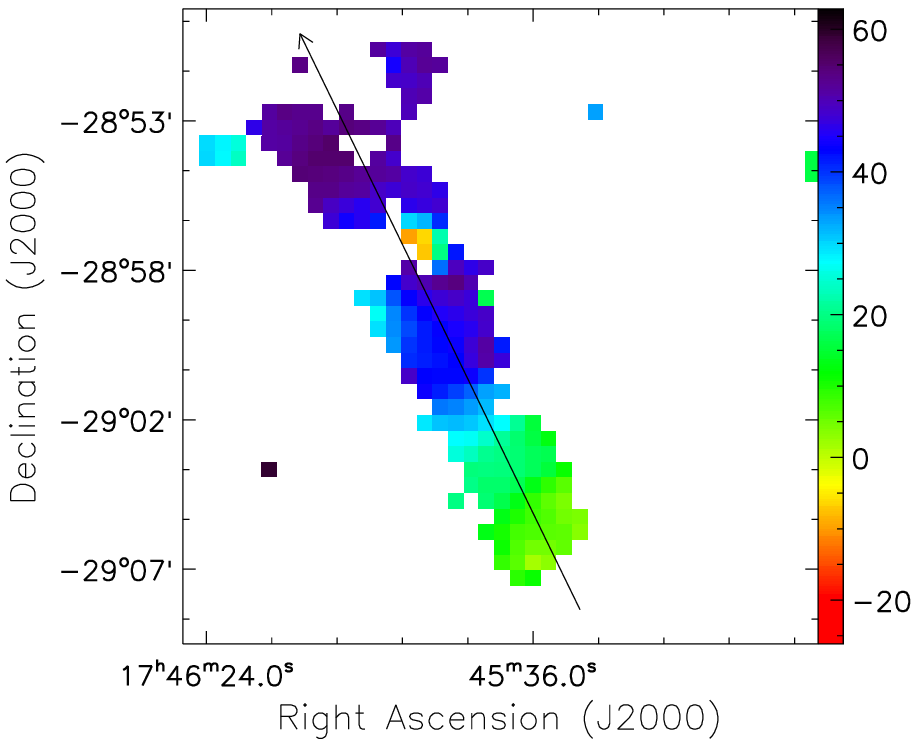}
\includegraphics[width=7.2cm,angle=0]{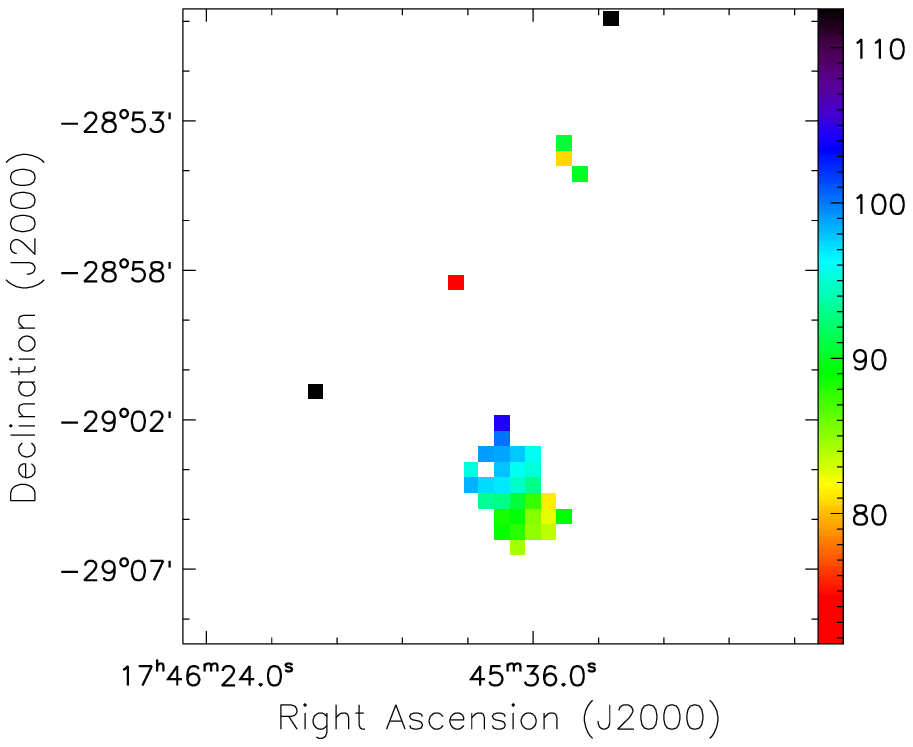}
\caption{Velocity maps of 95 GHz CH$_3$OH. Color bar units are km s$^{-1}$. Color images show the intensity weighted velocity (moment 1), $V_\mathrm{lsr}$, distribution of 95 GHz CH$_3$OH. The left and right panels show the velocity distribution of low (-30, 70) km s$^{-1}$ and high (70, 120) km s$^{-1}$ velocity components, respectively. The scales of the left and right panels are (-30, 70) and (70, 120) km s$^{-1}$, respectively. The long black arrow in the left panel indicates the axis of the position-velocity map shown in Figure 4.}
\end{figure*}

Figure 4 shows the position-velocity map along the black arrow marked in Figure 3. The peak emission is centered at $\sim 45$ km s$^{-1}$. There is also a variation of velocities from $\sim 10$ to $\sim$ 60 km s$^{-1}$ from the southwest to northeast, with corresponding velocity gradient $\sim$ 2 km s$^{-1}$ pc$^{-1}$. The velocity gradient decreases when the position passes $\sim$ 0\degr.20. There is a similar velocity gradient for the high velocity component (position ranges from $\sim$ 0\degr.02 to 0\degr.12), but the emission is much weaker.

\begin{figure*}[!ht]
\centering
\includegraphics[width=13cm,angle=0]{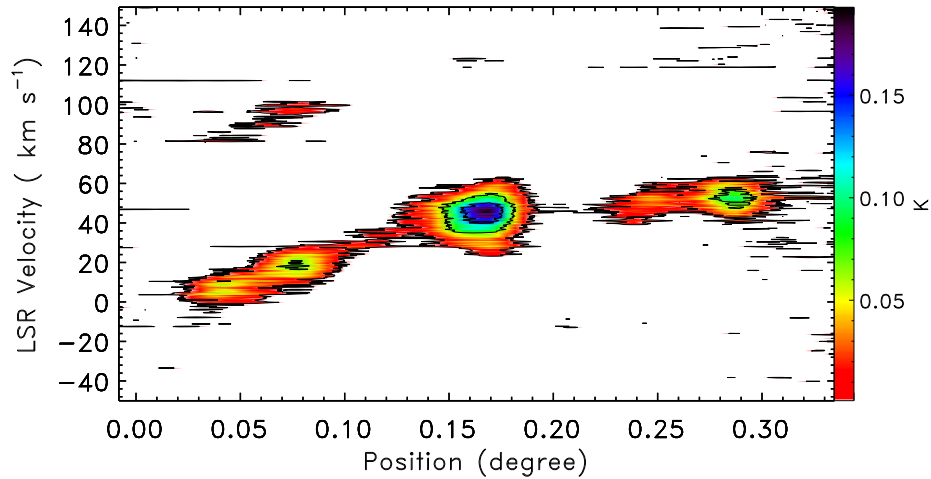}
\caption{Position-velocity map of 95 GHz CH$_3$OH along the arrow shown in Figure 3. Spectra at each position were averaged over 5$\arcmin$  perpendicular to the axis. 0\degr.00 on the x-axis corresponds to (R.A., Decl.)=(17$^{\mathrm{h}}$45$^{\mathrm{m}}$29.1$^{\mathrm{s}}$, -29$\degr$08$\arcmin$30$\arcsec$); and 0\degr.34 to (R.A., Decl.)=(17$^{\mathrm{h}}$46$^{\mathrm{m}}$10.3$^{\mathrm{s}}$, -28$\degr$50$\arcmin$00$\arcsec$). Contour levels are 15\%, 45\%, and 75\% of the maximum value (0.19 K).}
\end{figure*}


\section{Is 95 GHz \choh Emission Maser Emission?}

Similar to previous work toward the vicinities of Sgr A and Sgr B2 \citep{VES2000}, the broad line profiles in Figure 2 suggest that thermal emission contributes significantly to the detections.

The distribution of 95 GHz \choh emission is concentrated. In order to avoid the emission peaks being divided by pixels, the central zone is assume to contain four pixels, i.e., 1$\arcmin$ square. The central zone peak intensity is much higher than the pixels outside it. For peaks 1--5, integrated intensities outside the central zone are $\lesssim$ 60\% of the central zone peak intensity. This implies that peak intensities arise from compact areas, i.e., the emission has limited spatial extent, which indicates that maser emission may make a contribution to the total 95 GHz \choh emission in peaks 1--5.

The 95 GHz \choh maser originates in the same clump of gas that produces the 36 and 44 GHz \choh maser \citep{PM1990}. Figure 5 indicates that the 95 GHz \choh emission correlates with the 36/44 GHz \choh maser emission in some regions, and most of those co-location regions are around or in peaks 6 -- 8. This also indicates that part of 95 GHz \choh emission may be maser emission, particularly in regions around or in peaks 6 -- 8. The velocities of 36/44 GHz \choh masers in Figure 5 are less than 60 km s$^{-1}$, so we did not compare these with the high velocity component of 95 GHz \choh emission.

\begin{figure*}
\centering
\includegraphics[width=16cm,angle=0]{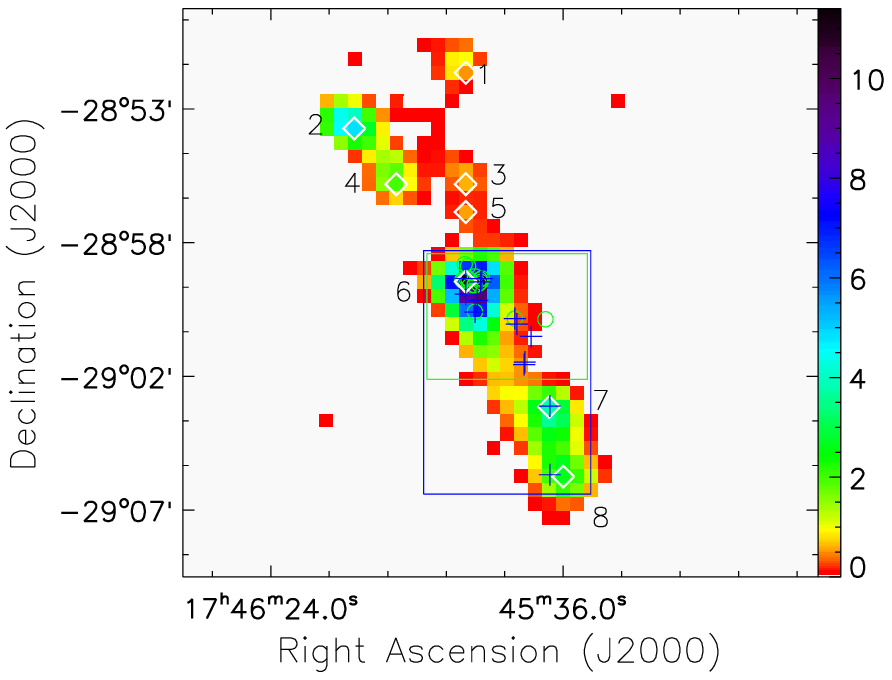}
\caption{Relative positions of the 36 GHz \choh masers \citep[blue pluses;][and references therein]{LW1996, SPF2010}, 44 GHz \choh masers \citep[green open circles; ][and references therein]{PSF2011}, and 95 GHz \choh emission (color image, integrated over -30 to 70 km s$^{-1}$, the scale is 10.6 K km$^{-1}$). Color bar units are K km s$^{-1}$. The blue and green rectangles indicate the study regions of 36 and 44 GHz \choh masers in those references, respectively.}
\end{figure*}

Thus, there may be a small fraction of 95 GHz \choh maser emission buried within thermal emission, and higher resolution and sensitivity observations would help to confirm this speculation.

\section{Excitation Source for 95 GHZ \choh Emission}

In this section, we extended peaks to clouds that were distinguishable from each other and clearly covered the corresponding peaks, and numbered clouds accordingly (e.g., cloud 1 corresponds to peak 1). We discussed the possible excitation sources for 95 GHz \choh emission. Surveys have revealed that class I methanol masers, such as 36 and 44 GHz methanol masers, are associated with not only star formation \citep{KJB2010, CEH2012}, but also SNRs \citep{LAV2011, PSF2011, PSF2014}. To investigate which one dominates the excitation for 95 GHz \choh emission in GCSNRR, we combined data of 20 cm continuum emission, 36 GHz \choh and 1720 MHz OH maser emission, HII regions, and WISE data. We also discussed the influence of tidal action, because GCSNRR is close to the Galactic center. We also discussed why 95 GHz \choh emission was only detected in GCSNRR. A summary was given in the end of this section.

\subsection{Interaction between 95 GHz \choh Emission and SNRs}

\subsubsection{Relative Position among 95 GHz \choh Emission and SNRs or Superbubble}

The 5 GHz and 20 cm continuum emission can trace both HII regions and SNRs in the vicinity of Sgr A East \citep{SLA1992, YHC2004}. For example, there are distinct differences in the 20 cm continuum emission map between HII regions \citep[][ Figure 19b, 21a and 23b]{YHC2004} and SNRs \citep[][ Figure 19b and 21a]{YHC2004}. We used the 20 cm continuum emission map to indicate SNRs. Figure 6 shows the 95 GHz \choh emission contours superimposed on the 20 cm radio continuum emission image from \citet{YHC2004}. To show the 20 cm radio continuum emission more explicitly, the two images have different maximal values: the left panel = 4.5 Jy beam$^{-1}$ and the right panel = 1.0 Jy beam$^{-1}$.

\begin{figure*}[!ht]
\centering
\includegraphics[width=7.2cm,angle=0]{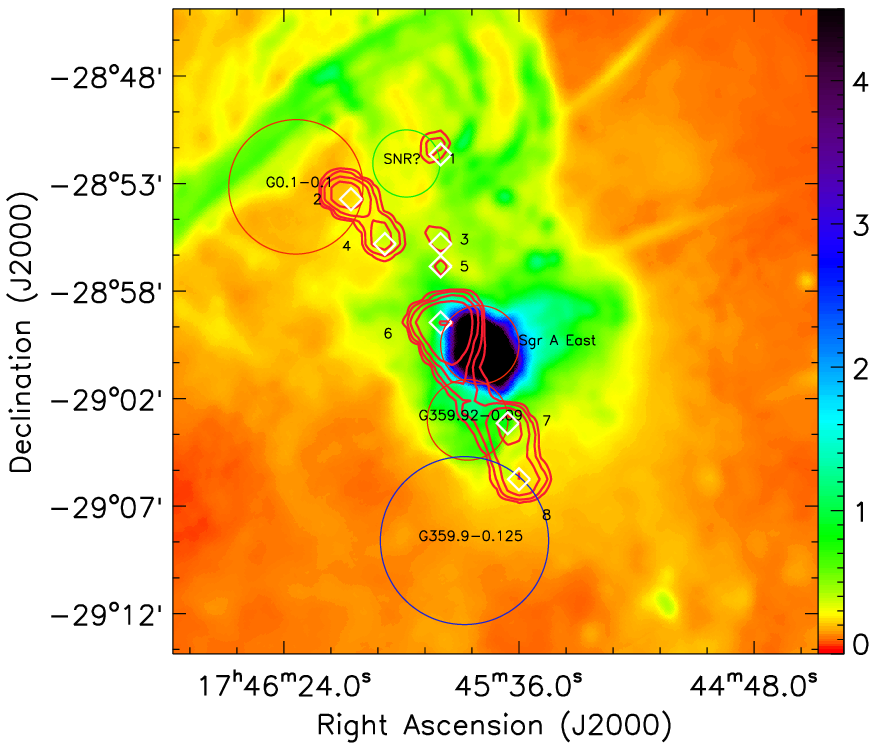}
\includegraphics[width=7.2cm,angle=0]{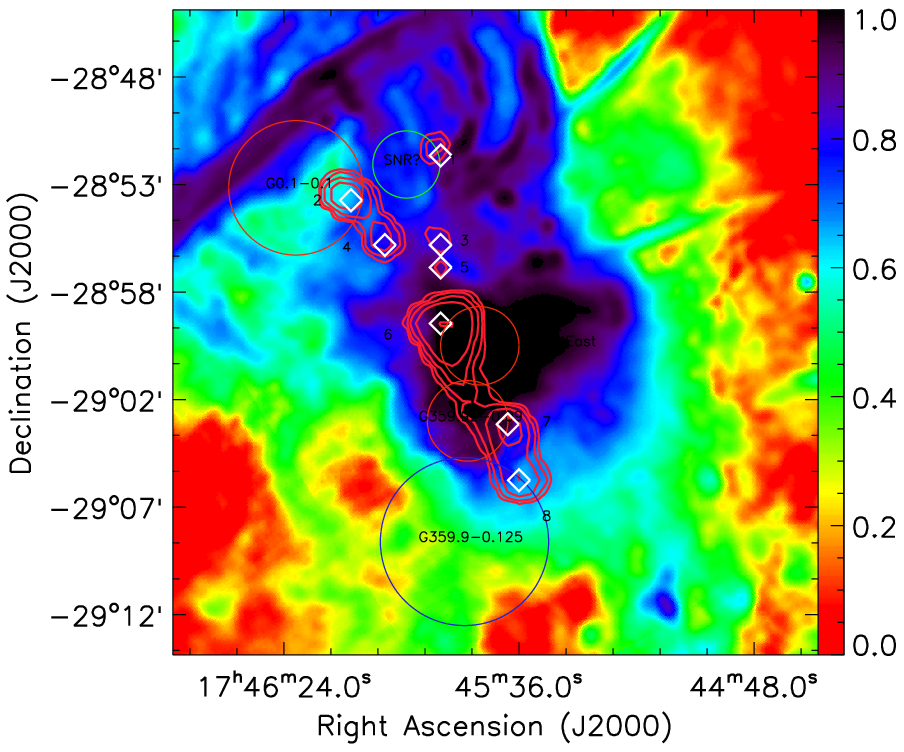}
\caption{Contours of 95 GHz \choh emission between -30 and 70 km s$^{-1}$ superimposed on the 20 cm radio continuum image from \citet{YHC2004}. Contours start at 3$\sigma$ and increase in steps of 1, 2, 4, 8, and 16$\times$3$\sigma$, where $\sigma=0.11$ K km s$^{-1}$. Color bars units are Jy beam$^{-1}$, and the scales of the left and right panels are 4.5 and 1.0 Jy beam$^{-1}$, respectively. The red circles indicate SNRs, the green circle indicates an SNR candidate (marked as ``SNR?''), and the blue circle indicates an superbubble candidate. The circles indicating G 359.92-0.09 and Sgr A East are the real angular size, and that of G 359.9-0.125 is half the real angular size. The sizes of SNR G 0.1-0.1 and ``SNR?'' are uncertain. The diamonds indicate peak positions, and contours outline clouds.}
\end{figure*}

The blue circle in figure 6 indicates a superbubble candidate, G 359.9-0.125, centered at $(l,b)\sim(359.84\degr, -0.14\degr)$ with size $15\arcmin\times3\arcmin$ \citep[][and references therein]{MTH2008, PMT2015}. The red circles indicate three SNRs: Sgr A East, G 0.1-0.1 and G 359.92-0.09. SNRs Sgr A East is clear in the left panel, and the right panel probably shows a shell-like structure in the vicinity of cloud 1. This shell-like structure may indicate an SNR candidate located at R.A. $\sim$ 17$^{\mathrm{h}}$46$^{\mathrm{m}}$00$^{\mathrm{s}}$, Decl. $\sim$ -28$\degr$52$\arcmin$12$\arcsec$; and represented by a green circle marked with ``SNR?''.

Cloud 1 lies in the northwest of the SNR candidate. Clouds 2 and 4 are in the southwest of the SNR G 0.1-0.1 and located in molecular cloud G 0.13-0.13. They are probably interacting with G 0.1-0.1 just as G 0.13-0.13 doing \citep{YLW2002}. The ridge connecting clouds 6, 7 and 8 \citep[i.e., Mol. Ridge or Molecular Ridge in][]{HH2005, AMM2011} surrounds the eastern edge of Sgr A East. SNR G 359.92-0.09, which is located to the east of them, probably has a strong interaction with Sgr A East and the eastern edge of the 20 km s$^{-1}$ cloud \citep{CH2000, HH2005, AMM2011}, where clouds 7 and 8 are located. The southeast of clouds 8 and 9 (the position of peak 9 overlaps with peak 8, see figure 1) may interact with superbubble candidate G 359.9-0.125, as shown in Figure 6. Most of the 95 GHz \choh emission surround SNRs (Figure 6), which suggests that the emission may be excited by interaction with SNRs.

\subsubsection{Line Widths of 95 GHz \choh Emission Versus SNRs or Superbubble}

Molecular line widths can be broadened by interaction between shock and molecular cloud \citep{KBR2014}. Furthermore, line broadenings were regarded as strong kinematic evidence for the interaction between SNR and molecular cloud \citep[][and references therein]{JCW2010, KBR2016}. For instance, the broad CS (1-0) line widths were used to indicate the interaction between the Sgr A East shell and the 50 km s$^{-1}$ molecular cloud \citep{TMO2009}. Therefore, 95 GHz \choh line width broadening may be an indicator of interaction among SNRs and 95 GHz \choh clouds. Figure 7 shows the distribution of line widths for 95 GHz \choh emission. The line for the high velocity component is randomly distributed in the region around peak 9, and therefore omitted in Figure 7.

\begin{figure*}[!ht]
\vspace{-2cm}
\centering
\includegraphics[width=16cm,angle=0]{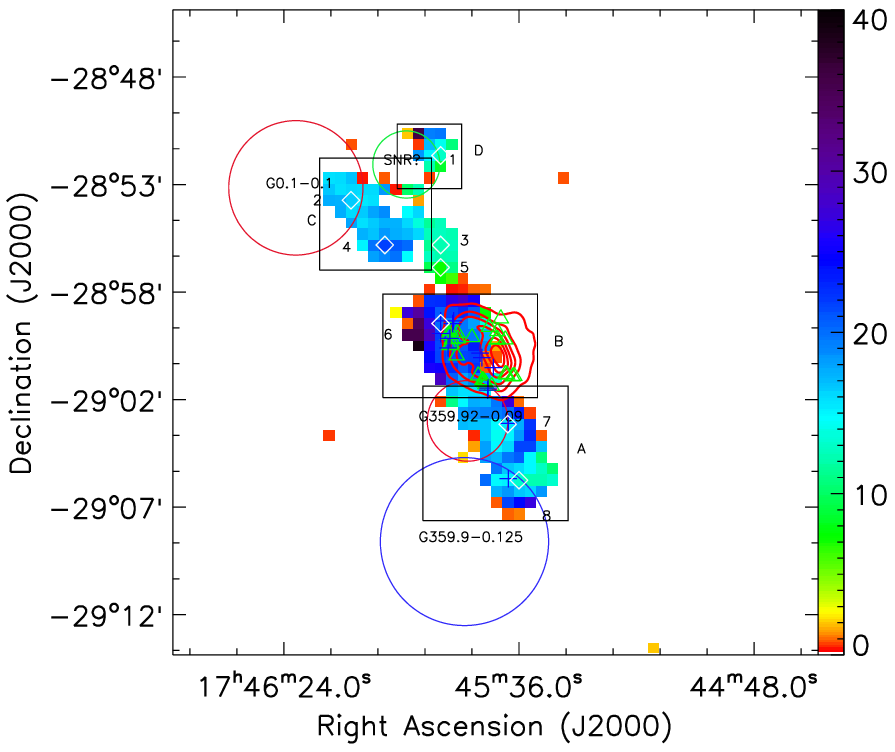}
\caption{Distribution of 95 GHz \choh line widths (color image). Maximum is 38.54 km s$^{-1}$. Color bar units are km s$^{-1}$. Red contours of the 20 cm continuum image \citep{YHC2004} show Sgr A East. The lowest contour level and contour interval are 15\% of the maximum (11.23 Jy beam$^{-1}$). Green triangles indicate 1720 MHz OH masers \citep[see][and references therein]{SP2008}, and blue pluses indicate 36 GHz \choh masers \citep[and references therein]{LW1996, SPF2010}. The four rectangles present four interesting regions.}
\end{figure*}

Figure 7 shows four interesting regions marked by rectangles:

\textbf{Region A.} Line widths relatively higher than their neighborhoods show a reversed-S structure crossing the interface between clouds 7 and 8. In this region, the north part of the reversed-S structure is located west of cloud 7, possibly indicating an interaction among cloud 7, SNR G 359.92-0.09, and Sgr A East. The midsection of the reversed-S structure may indicate SNR G 359.92-0.09 rushing toward cloud 8, or may trace an interaction between clouds 7 and 8. The south part of the reversed-S structure lies south of cloud 8, which may indicate an interaction between cloud 8 and superbubble candidate G 359.9-0.125.

\textbf{Region B.} Line widths in cloud 6 (typical value is $\sim$20 km s$^{-1}$) are much higher than in other regions, and increase toward the east. This possibly indicates that the 50 km s$^{-1}$ cloud (including cloud 6) has a strong interaction with Sgr A East.

\textbf{Region C.} Line widths increase toward the southeast of cloud 4 and west of cloud 2, which constitutes a faint shell. This may suggest an interaction between the 95 GHz \choh cloud and SNR G 0.1-0.1.

\textbf{Region D.} Line widths are higher in the northeast of cloud 1, which may indicate that cloud 1 is interacting with an SNR candidate as discussed above to the east of cloud 1 (see the green circle in the right panel of Figure 6 and Figure 7). This emission may be also excited by star formation as discussed in Section 5.2.

Figure 7 also shows that 95 GHz CH$_3$OH line widths are associated with the 36 GHz \choh masers in the south of Decl. $\sim$ -28$^\circ$58$\arcmin$ and 1720 MHz OH masers around Sgr A East. The geometry of the 36 GHz masers in Sgr A East outlines the current location of an SNR shock front \citep{SP2012}. Most 1720 MHz OH masers arise in regions where the SNR Sgr A East is interacting with the 20 and 50 km s$^{-1}$ molecular clouds and the nearby SNR G 359.92-0.09 \citep{SP2008}. The 50 km s$^{-1}$ molecular cloud includes cloud 6, and the 20 km s$^{-1}$ molecular cloud includes clouds 7 and 8.

As a summary of this subsection, 95 GHz \choh emission may be excited by interaction with SNRs and its line width is probably a good tracer of SNRs.

\subsection{Effect of Star Formation}

The 95 GHz \choh emission has usually assumed to be associated with star formation \citep{CES2011, YXC2017}. HII regions and WISE data were used to indicate star formation regions in this study. Figure 8 superimposes the positions of HII regions \citep[yellow crosses, abstracted from][and reference therein]{GSG1985, KC1997, PBD2003, ABB2014} on the WISE false color map, along with the 95 GHz \choh emission contours. Both the HII regions and the WISE emission show that 95 GHz \choh emission deviates slightly from the star formation regions.

\begin{figure*}[!ht]
\centering
\includegraphics[width=12.468cm,height=9.84cm,angle=0]{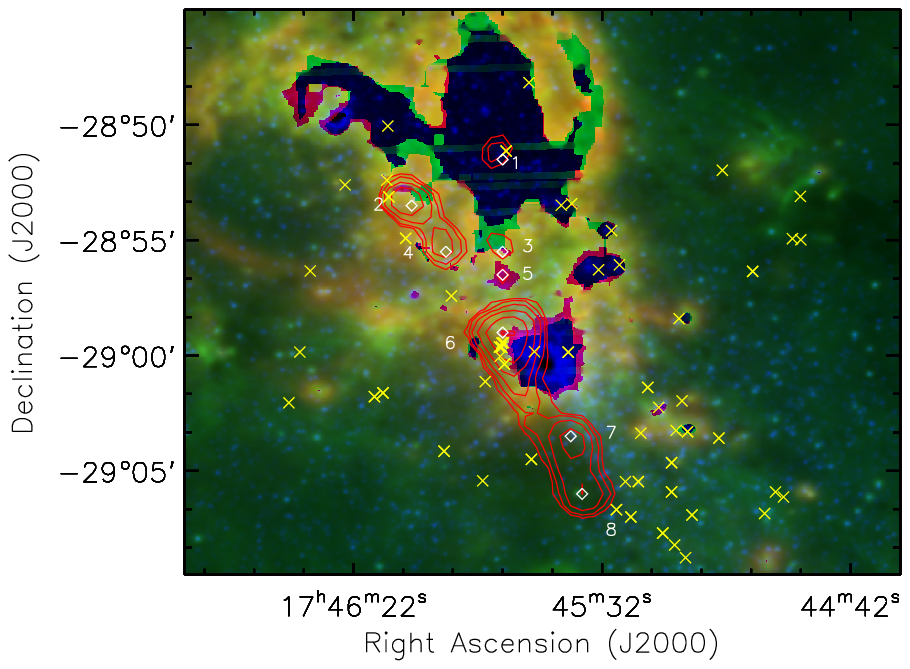}
\caption{Background false color map, where blue is the WISE 4.6 $\mu$m, green is 12 $\mu$m, and red is 22 $\mu$m data, with 95 GHz \choh emission contours between -30 and 70 km s$^{-1}$ and HII regions superimposed \citep[yellow crosses, abstracted from][and references therein]{GSG1985, KC1997, PBD2003, ABB2014}. Contours start at 3$\sigma$ and increase in steps of 1, 2, 4, 8, and 16$\times$3$\sigma$, where $\sigma=0.11$ K km s$^{-1}$. The diamonds indicate peak positions, and contours outline clouds.}
\end{figure*}

The region, where star formation is most likely to be the dominant excitation source for 95 GHz \choh emission in GCSNRR, is cloud 1, because the nearest HII region lies just 0.40$\arcmin$ west of peak 1. The WISE emission also shows that star formation is probably play an important role in exciting \choh emission in clouds 2 -- 5.

Star formation may not be the dominant factor to excite 95 GHz \choh emission in the 50 km s$^{-1}$ cloud (including cloud 6), because the energy restored in star formation process may be less than the kinetic energy in 50 km s$^{-1}$ cloud. Specifically, the energy restored in protostellar outflows is up to $10^{47}$ -- $10^{48}$ erg \citep{B1996, L2000}. That is far less than the kinetic energy in the 50 km s$^{-1}$ cloud ($0.5m\cdot\Delta V^2\sim10^{50}$ erg), where its mass, $m$, is $\sim$5$\times10^4\;\mathrm{M}_{\sun}$ \citep{HH2005} and line width, $\Delta V$, in terms of CH$_3$OH $\sim 20$ km s$^{-1}$ (no less than 10 km s$^{-1}$). Therefore, SNR Sgr A East is more likely to be the dominant excitation source than star formation in the 50 km s$^{-1}$ cloud (including cloud 6).

In clouds 7 -- 9, both HII regions and WISE emission deviate from 95 GHz \choh emission, which suggests that the star formation action is probably not the dominant excitation source in clouds 7 -- 9.

\subsection{Influence of Tidal Action}

The Roche density $n_{\mathrm{RL}}$ is given by \citep{MG2016}:

\begin{equation}
 n_{\mathrm{RL}}\sim 10^7\;\mathrm{cm}^{-3}\left(\frac{m_{\mathrm{BH}}}{3\times 10^6\;\mathrm{M}_{\sun}}\right)\left(\frac{\mathrm{pc}}{r}\right)^3,
\end{equation}

\noindent where $m_{\mathrm{BH}}=4.30\pm0.20_{\mathrm{stat}}\pm0.30_{\mathrm{sys}}\times 10^6\;\mathrm{M}_{\sun}$  is the mass of the supermassive black hole \citep{GEF2009, MG2016}, and $r$ is the distance of the molecular cloud from the supermassive black hole. Clouds 6 -- 8 are located in the 20 or 50 km s$^{-1}$ clouds, which are $\sim 20$ pc from the supermassive black hole Sgr A$^{\ast}$ \citep{MG2016}. Then $n_{\mathrm{RL}}$ is no less than $1.8\times 10^6$ cm$^{-3}$, which is much higher than the number densities in 20 and 50 km s$^{-1}$ clouds \citep[$0.4$ -- $4\times10^5$ cm$^{-3}$, see Table 3 of][]{AMM2011}. This indicates that the 20 and 50 km s$^{-1}$ clouds are currently being tidally disrupted according to \citet{MG2016}, which is also suggested by \citet{HH2005}.

Assuming near equatorial and near-circular orbit about Sgr A$^{\ast}$, the tidal potential, $V_{\mathrm{tidal}}$, can be estimated as \citep{MD2000}

\begin{equation}
 V_{\mathrm{tidal}}=\frac{Gm_{\mathrm{BH}}}{4a^3}R^2(3\cos\psi+1),
\end{equation}

\noindent where $a$ and $R$ are the distance of the 50 km s$^{-1}$ cloud from the black hole, and the radius of the black hole, respectively, and $\psi$ is the angle between the projected radius vector and the plane perpendicular to the link line between the center of the 50 km s$^{-1}$ cloud and the black hole. Even if $a=R=0.04$ pc, where 0.04 pc is the upper limit of the radius of the black hole \citep{AMM2011}, $V_{\mathrm{tidal}}\sim 10^{18}$ erg, which is far less than the kinetic energy in 50 km s$^{-1}$ cloud (i.e., $\sim10^{50}$ erg, see section 5.2). That indicates that tidal action is not likely to be a source to excite \choh emission, although the estimate above is quite approximate. Similar conclusion applies to the 20 km s$^{-1}$ cloud.

\subsection{The Role of Cosmic Rays}

Observing 36 GHz \choh in CMZ (covers GCSNRR), \citet{YCW2013} explained the enhanced abundance of \choh in terms of interactions between cosmic rays and molecular gas. Since 95 GHz \choh emission is associated with 36 GHz \choh emission (see Figure 5 and Figure 7), the 95 GHz \choh emission can be enhanced by these interactions in CMZ.

\subsection{The Possible Reason Why 95 GHz \choh Emission Was Only Detected in GCSNRR}

Table 1 lists the properties of the eight SNRs, and in particularly indicates that the type of SNRs is the unique character of SNRs Sgr A East and G 0.1-0.1 with the detection of 95 GHz \choh emission. SNR G 0.1-0.1 and Sgr A East are probably thermal \& plerionic composite \citep{HW2013b} and thermal composite \citep{YWR2003, V2012} SNRs, respectively.

Thermal composite SNRs have interior thermal X-ray emission, and are also known as mixed morphology SNRs (MM SNRs). \citet{FGR1996} and \citet{GFG1997} first noted that the OH (1720 MHz) SNRs (SNRs that have detected 1720 MHz OH masers) belonged predominantly to a particular class of MM SNRs that have center-filled thermal X-ray emission. This hypothesis was verified on a firmer statistical footing by \citet{YWR2003}. Interactions between SNR and molecular cloud play an important role in producing this interior X-ray gas \citep{F2011}. The 95 GHz \choh emission in cloud 6 is associated with 1720 MHz OH maser emission, which also supports a possible correlation between SNR type and 95 GHz \choh emission. This correlation could also be one of the possible reasons why 95 GHz \choh emission was only detected in GCSNRR.

On the other side, interaction between cosmic rays and molecular gas may be another reason why 95 GHz \choh emission was only detected in GCSNRR. Cosmic rays collide with ambient gas produce $\gamma$-rays \citep{F2011}, such as GeV/TeV emission. The region surrounding the Galactic center is among the brightest and most complex in high-energy $\gamma$-rays \citep[][and references therein]{AAlA2016}. Although both the GeV and TeV fluxes around Sgr A East are less than that in G 184.6-5.8, they are much higher than that in G 1.4-0.1, G 29.7-0.3, G 111.7-2.1 and G 120.1+1.4 (see Table 1). In addition, it is unlikely that SNR G 184.6-5.8 interacts with molecular clouds \citep{GWL1990, FSP2007, KBR2016}. Therefore, the interaction between cosmic rays and molecular gas near the Galactic center could still be one of the possible reasons why 95 GHz \choh emission was only detected in GCSNRR.

As a summary of this section, 95 GHz \choh emission is probably excited by interaction with SNRs and may also correlate with the type of SNRs. Although star formation may play an important role in exciting \choh emission 95 GHz \choh emission in some regions, it is probably not the dominant excitation source in clouds 7 -- 9, and may not be the dominant excitation source in cloud 6. It is difficult to clarify whether star formation or SNRs is the dominant excitation source in clouds 1 -- 5. Interaction between cosmic rays and molecular gas in CMZ may enhance 95 GHz \choh emission. The contribution from tidal action to exciting 95 GHz \choh emission is negligible. We also suggest that the possible reasons why 95 GHz \choh emission was only detected in GCSNRR may be correlated with the type of SNRs near the Galactic center, and the interaction between cosmic rays and molecular gas in CMZ. However, a larger sample in the future survey is required to confirm the correlation between 95 GHz \choh emission and the type of SNRs.

\section{Summary}

We conducted a survey for 95 GHz ($8_0-7_1$ A$^{+}$) \choh emission toward eight SNRs with angular size $\lesssim$ 10$\arcmin$. The main summaries are as follows:

\begin{enumerate}
 \item 95 GHz \choh emission was only detected in Sgr A East, G 0.1-0.1 and G 359.92-0.09. The emission can be decomposed into nine spatial peaks with velocity range of eight peaks being (-30, 70) km s$^{-1}$, and the other (70, 120) km s$^{-1}$.
 \item Part of the 95 GHz \choh emission may be maser emission in some regions, particularly in regions around or in peaks 1 -- 8. Higher resolution and sensitivity observations are required to confirm this speculation.
 \item Most 95 GHz \choh emission surrounds SNRs, and is probably excited by interacting with SNRs in CMZ, although star formation is probably important to excite 95 GHz \choh emission in some regions of CMZ. The influence of tidal action is negligible.
 \item SNR type and interaction between cosmic rays and molecular gas in CMZ could be possible reasons why 95 GHz \choh emission was only detected toward SNRs near the Galactic center. A lager sample is required to confirm the correlation between 95 GHz \choh emission and the type of SNRs.
\end{enumerate}

\normalem
\begin{acknowledgements}
We acknowledge Fa-cheng Li, Yuliang Xin, Zhaoqiang Shen and Yuehui Ma for their valuable help, and thank Farhad Yusef-Zadeh for providing 20 cm radio image. This work was supported by the National Science Foundation of China (Grant Numbers: 11673066, 11233007, 11590781 and 11273043), and the Key Laboratory for Radio Astronomy.
\end{acknowledgements}


\begin{thebibliography}{99}
\bibitem[Acciari et al.(2011)]{AAA2011} Acciari, V.~A., Aliu, E., Arlen, T., et al.\ 2011, \apjl, 730, L20
\bibitem[Acero et al.(2016)]{AAA2016} Acero, F., Ackermann, M., Ajello, M., et al.\ 2016, \apjs, 224, 8
\bibitem[Aharonian et al.(2001)]{AAB2001} Aharonian, F., Akhperjanian, A., Barrio, J., et al.\ 2001, \aap, 370, 112
\bibitem[Aharonian et al.(2008)]{AAB2008} Aharonian, F., Akhperjanian, A.~G., Barres de Almeida, U., et al.\ 2008, \aap, 488, 219
\bibitem[Aharonian et al.(2006)]{AABe2006} Aharonian, F., Akhperjanian, A.~G., Bazer-Bachi, A.~R., et al.\ 2006, \aap, 457, 899
\bibitem[Ajello et al.(2016)]{AAlA2016} Ajello, M., Albert, A., Atwood, W.~B., et al.\ 2016, \apj, 819, 44
\bibitem[Albert et al.(2007)]{AAA2007} Albert, J., Aliu, E., Anderhub, H., et al.\ 2007, \aap, 474, 937
\bibitem[Amo-Baladr{\'o}n et al.(2011)]{AMM2011} Amo-Baladr{\'o}n, M.~A., Mart{\'{\i}}n-Pintado, J., \& Mart{\'{\i}}n, S.\ 2011, \aap, 526, A54
\bibitem[Anderson et al.(2014)]{ABB2014} Anderson, L.~D., Bania, T.~M., Balser, D.~S., et al.\ 2014, \apjs, 212, 1
\bibitem[Bachiller(1996)]{B1996} Bachiller, R.\ 1996, \araa, 34, 111
\bibitem[Ball et al.(1970)]{BGL1970} Ball, J.~A., Gottlieb, C.~A., Lilley, A.~E., \& Radford, H.~E.\ 1970, \apjl, 162, L203
\bibitem[Becker et al.(1983)]{BHS1983} Becker, R.~H., Helfand, D.~J., \& Szymkowiak, A.~E.\ 1983, \apjl, 268, L93
\bibitem[Belloche et al.(2013)]{BMM2013} Belloche, A., M{\"u}ller, H.~S.~P., Menten, K.~M., Schilke, P., \& Comito, C.\ 2013, \aap, 559, A47
\bibitem[Blanton \& Helfand(1996)]{BH1996} Blanton, E.~L., \& Helfand, D.~J.\ 1996, \apj, 470, 961
\bibitem[Bochow(2011)]{B2011} Bochow, A.\ 2011, Ph.D.~Thesis,
\bibitem[Chen et al.(2012)]{CEH2012} Chen, X., Ellingsen, S.~P., He, J.-H., et al.\ 2012, \apjs, 200, 5
\bibitem[Chen et al.(2011)]{CES2011} Chen, X., Ellingsen, S.~P., Shen, Z.-Q., Titmarsh, A., \& Gan, C.-G.\ 2011, \apjs, 196, 9
\bibitem[Coil \& Ho(2000)]{CH2000} Coil, A.~L., \& Ho, P.~T.~P.\ 2000, \apj, 533, 245
\bibitem[Davies et al.(2009)]{DOK2009} Davies, B., Origlia, L., Kudritzki, R.-P., et al.\ 2009, \apj, 694, 46
\bibitem[Ferrand \& Safi-Harb(2012)]{FS2012} Ferrand, G., \& Safi-Harb, S.\ 2012, Advances in Space Research, 49, 1313
\bibitem[Fish et al.(2007)]{FSP2007} Fish, V.~L., Sjouwerman, L.~O., \& Pihlstr{\"o}m, Y.~M.\ 2007, \apjl, 670, L117
\bibitem[Frail(2011)]{F2011} Frail, D.~A.\ 2011, \memsai, 82, 703
\bibitem[Frail et al.(1996)]{FGR1996} Frail, D.~A., Goss, W.~M., Reynoso, E.~M., et al.\ 1996, \aj, 111, 1651
\bibitem[Gillessen et al.(2009)]{GEF2009} Gillessen, S., Eisenhauer, F., Fritz, T.~K., et al.\ 2009, \apjl, 707, L114
\bibitem[Goss et al.(1985)]{GSG1985} Goss, W.~M., Schwarz, U.~J., van Gorkom, J.~H., \& Ekers, R.~D.\ 1985, \mnras, 215, 69P
\bibitem[Graham et al.(1990)]{GWL1990} Graham, J.~R., Wright, G.~S., \& Longmore, A.~J.\ 1990, \apj, 352, 172
\bibitem[Green(2014)]{G2014} Green, D.~A.\ 2014, Bulletin of the Astronomical Society of India, 42, 47
\bibitem[Green et al.(1997)]{GFG1997} Green, A.~J., Frail, D.~A., Goss, W.~M., \& Otrupcek, R.\ 1997, \aj, 114, 2058
\bibitem[Hatchell et al.(1998)]{HTM1998} Hatchell, J., Thompson, M.~A., Millar, T.~J., \& MacDonald, G.~H.\ 1998, \aaps, 133, 29
\bibitem[Heard \& Warwick(2013)]{HW2013b} Heard, V., \& Warwick, R.~S.\ 2013B, \mnras, 434, 1339
\bibitem[Herrnstein \& Ho(2005)]{HH2005} Herrnstein, R.~M., \& Ho, P.~T.~P.\ 2005, \apj, 620, 287
\bibitem[Jiang et al.(2010)]{JCW2010} Jiang, B., Chen, Y., Wang, J., et al.\ 2010, \apj, 712, 1147
\bibitem[Kalenskii et al.(1997)]{KDB1997} Kalenskii, S.~V., Dzura, A.~M., Booth, R.~S., Winnberg, A., \& Alakoz, A.~V.\ 1997, \aap, 321, 311
\bibitem[Kalenski{\u i} et al.(2006)]{KPS2006} Kalenski{\u i}, S.~V., Promyslov, V.~G., Slysh, V.~I., Bergman, P., \& Winnberg, A.\ 2006, Astronomy Reports, 50, 289
\bibitem[Kalenskii et al.(2010)]{KJB2010} Kalenskii, S.~V., Johansson, L.~E.~B., Bergman, P., et al.\ 2010, \mnras, 405, 613
\bibitem[Kilpatrick et al.(2014)]{KBR2014} Kilpatrick, C.~D., Bieging, J.~H., \& Rieke, G.~H.\ 2014, \apj, 796, 144
\bibitem[Kilpatrick et al.(2016)]{KBR2016} Kilpatrick, C.~D., Bieging, J.~H., \& Rieke, G.~H.\ 2016, \apj, 816, 1
\bibitem[Koralesky et al.(1998)]{KFG1998} Koralesky, B., Frail, D.~A., Goss, W.~M., Claussen, M.~J., \& Green, A.~J.\ 1998, \aj, 116, 1323
\bibitem[Kuchar \& Clark(1997)]{KC1997} Kuchar, T.~A., \& Clark, F.~O.\ 1997, \apj, 488, 224
\bibitem[Leurini et al.(2016)]{LMW2016} Leurini, S., Menten, K.~M., \& Walmsley, C.~M.\ 2016, \aap, 592, A31
\bibitem[Leurini et al.(2004)]{LSM2004} Leurini, S., Schilke, P., Menten, K.~M., et al.\ 2004, \aap, 422, 573
\bibitem[Liechti \& Wilson(1996)]{LW1996} Liechti, S., \& Wilson, T.~L.\ 1996, \aap, 314, 615
\bibitem[Litovchenko et al.(2011)]{LAV2011} Litovchenko, I.~D., Alakoz, A.~V., Val'Tts, I.~E., \& Larionov, G.~M.\ 2011, Astronomy Reports, 55, 978
\bibitem[Livio(2000)]{L2000} Livio, M.\ 2000, Unsolved Problems in Stellar Evolution,
\bibitem[Maeda et al.(2002)]{MBF2002} Maeda, Y., Baganoff, F.~K., Feigelson, E.~D., et al.\ 2002, \apj, 570, 671
\bibitem[Mapelli \& Gualandris(2016)]{MG2016} Mapelli, M., \& Gualandris, A.\ 2016, Lecture Notes in Physics, Berlin Springer Verlag, 905, 205
\bibitem[McEwen et al.(2014)]{MPS2014} McEwen, B.~C., Pihlstr{\"o}m, Y.~M., \& Sjouwerman, L.~O.\ 2014, \apj, 793, 133
\bibitem[Menten et al.(1988a)]{MWH1988} Menten, K.~M., Walmsley, C.~M., Henkel, C., \& Wilson, T.~L.\ 1988, \aap, 198, 253
\bibitem[Menten et al.(1988b)]{MRM1988} Menten, K.~M., Reid, M.~J., Moran, J.~M., et al.\ 1988, \apjl, 333, L83
\bibitem[Mori et al.(2008)]{MTH2008} Mori, H., Tsuru, T.~G., Hyodo, Y., Koyama, K., \& Senda, A.\ 2008, \pasj, 60, 183
\bibitem[Murray \& Dermott(2000)]{MD2000} Murray, C.~D., \& Dermott, S.~F.\ 2000, Solar System Dynamics, by C.D.~Murray and S.F.~Dermott.~ISBN 0521575974.~Cambridge, UK: Cambridge University Press, 2000.
\bibitem[Nesterenok(2016)]{N2016} Nesterenok, A.~V.\ 2016, \mnras, 455, 3978
\bibitem[Paladini et al.(2003)]{PBD2003} Paladini, R., Burigana, C., Davies, R.~D., et al.\ 2003, \aap, 397, 213
\bibitem[Penzias \& Burrus (1973)]{PB1973}Penzias, A. A. \& Burrus, C. A. 1973, ARA\&A, 11, 51
\bibitem[Pihlstr{\"o}m et al.(2011)]{PSF2011} Pihlstr{\"o}m, Y.~M., Sjouwerman, L.~O., \& Fish, V.~L.\ 2011, \apjl, 739, L21
\bibitem[Pihlstr{\"o}m et al.(2014)]{PSF2014} Pihlstr{\"o}m, Y.~M., Sjouwerman, L.~O., Frail, D.~A., et al.\ 2014, \aj, 147, 73
\bibitem[Plambeck \& Menten(1990)]{PM1990} Plambeck, R.~L., \& Menten, K.~M.\ 1990, \apj, 364, 555
\bibitem[Ponti et al.(2015)]{PMT2015} Ponti, G., Morris, M.~R., Terrier, R., et al.\ 2015, \mnras, 453, 172
\bibitem[Predehl \& Schmitt(1995)]{PS1995} Predehl, P., \& Schmitt, J.~H.~M.~M.\ 1995, \aap, 293, 889
\bibitem[Serabyn \etal(1992)]{SLA1992} Serabyn, E., Lacy, J.~H., \& Achtermann, J.~M.\ 1992, \apj, 395, 166
\bibitem[Shan, Yang \& Shi (2012)]{YS2012}Shan, W., Yang, J., Shi, S., et al. 2012, IEEE Transactions on Terahertz Science and Technology, 2, 593
\bibitem[Sjouwerman \& Pihlstr{\"o}m(2008)]{SP2008} Sjouwerman, L.~O., \& Pihlstr{\"o}m, Y.~M.\ 2008, \apj, 681, 1287
\bibitem[Sjouwerman \& Pihlstr{\"o}m(2012)]{SP2012} Sjouwerman, L.~O., \& Pihlstr{\"o}m, Y.~M.\ 2012, IAU Symposium, 287, 449
\bibitem[Sjouwerman \etal(2010)]{SPF2010} Sjouwerman, L.~O., Pihlstr{\"o}m, Y.~M., \& Fish, V.~L.\ 2010, \apjl, 710, L111
\bibitem[Slysh et al.(1994)]{SBB1994} Slysh, V.~I., Bachiller, R., Berulis, I.~I., et al.\ 1994, \azh, 71, 37
\bibitem[Sugizaki et al.(2001)]{SMK2001} Sugizaki, M., Mitsuda, K., Kaneda, H., et al.\ 2001, \apjs, 134, 77
\bibitem[Tsuboi et al.(2009)]{TMO2009} Tsuboi, M., Miyazaki, A., \& Okumura, S.~K.\ 2009, \pasj, 61, 29
\bibitem[Ulich \& Haas (1976)]{UH1976}Ulich, B. L. \& Haas, R. W. 1976, ApJS, 30, 247
\bibitem[Val'tts et al.(2000)]{VES2000} Val'tts, I.~E., Ellingsen, S.~P., Slysh, V.~I., et al.\ 2000, \mnras, 317, 315
\bibitem[Vink(2012)]{V2012} Vink, J.\ 2012, \aapr, 20, 49
\bibitem[Wallace et al.(1999)]{WLK1999} Wallace, B.~J., Landecker, T.~L., Kalberla, P.~M.~W., \& Taylor, A.~R.\ 1999, \apjs, 124, 181
\bibitem[Wallace et al.(1994)]{WLT1994} Wallace, B.~J., Landecker, T.~L., \& Taylor, A.~R.\ 1994, \aap, 286, 565
\bibitem[Yang et al.(2017)]{YXC2017} Yang, W., Xu, Y., Chen, X., et al.\ 2017, arXiv:1705.01806
\bibitem[Yusef-Zadeh et al.(2013)]{YCW2013} Yusef-Zadeh, F., Cotton, W., Viti, S., Wardle, M., \& Royster, M.\ 2013, \apjl, 764, L19
\bibitem[Yusef-Zadeh et al.(1999)]{YGR1999} Yusef-Zadeh, F., Goss, W.~M., Roberts, D.~A., Robinson, B., \& Frail, D.~A.\ 1999, \apj, 527, 172
\bibitem[Yusef-Zadeh et al.(2004)]{YHC2004} Yusef-Zadeh, F., Hewitt, J.~W., \& Cotton, W.\ 2004, \apjs, 155, 421
\bibitem[Yusef-Zadeh et al.(2002a)]{YLW2002} Yusef-Zadeh, F., Law, C., \& Wardle, M.\ 2002A, \apjl, 568, L121
\bibitem[Yusef-Zadeh et al.(2002b)]{YLWe2002} Yusef-Zadeh, F., Law, C., Wardle, M., et al.\ 2002B, \apj, 570, 665
\bibitem[Yusef-Zadeh et al.(2003)]{YWR2003} Yusef-Zadeh, F., Wardle, M., Rho, J., \& Sakano, M.\ 2003, \apj, 585, 319
\bibitem[Zubrin \& Shulga(2008)]{ZS2008} Zubrin, S.~Y., \& Shulga, V.~M.\ 2008, Young Scientists 15th Proceedings, 41
\end{thebibliography}

\end{document}